\documentclass[useAMS,usenatbib,usegraphicx]{mn2e}

\def\ltsima{$\; \buildrel < \over \sim \;$}
\def\simlt{\lower.5ex\hbox{\ltsima}}
\def\gtsima{$\; \buildrel > \over \sim \;$}
\def\simgt{\lower.5ex\hbox{\gtsima}}
\def\gsimeq
{\hbox{\raise0.5ex\hbox{$>\lower1.06ex\hbox{$\kern-1.07em{\sim}$}$}}}
\def\lsimeq
{\hbox{\raise0.5ex\hbox{$<\lower1.06ex\hbox{$\kern-1.07em{\sim}$}$}}}

\def\xmm{{\it XMM-Newton }}

\def\sax{{\it BeppoSAX}}
\def\xmm{{\it XMM-Newton}}
\def\chandra{{\it Chandra}}
\def\suzaku{{\it Suzaku}}

\def\apj{ApJ}
\def\mnras{MNRAS}
\def\aap{A\&A}
\def\apjl{ApJ}
\def\apjs{ApJS}
\def\araa{ARA\&A}
\def\pasj{PASJ}

\def\xis{XIS}
\def\xis1{XIS1}
\def\xis2{XIS2}
\def\xis3{XIS3}

\title[] 
 {{XMM-Newton and Suzaku analysis of the Fe K complex in the Seyfert 1 galaxy Mrk~509}}

 \author[G.\ Ponti et al. ]
 {G.~Ponti$^{1,2,3}$\thanks{ponti@iasfbo.inaf.it},
   M. Cappi$^{2}$, C. Vignali$^{3}$, G. Miniutti$^{1,4}$, F. Tombesi$^{2,3}$, 
   M. Dadina$^{2,3}$, 
\newauthor
   A.C. Fabian$^{5}$, P. Grandi$^{2}$, J. Kaastra$^{6,7}$, 
   P.O. Petrucci$^{8}$, S. Bianchi$^{9}$, G. Matt$^{9}$, 
\newauthor
   L. Maraschi$^{10}$ 
   and G. Malaguti$^{2}$ 
\\ \\
   $^1$APC Universit\'e Paris 7 Denis Diderot, 75205 Paris Cedex 13, France \\
   $^2$INAF--IASF Bologna, Via Gobetti 101, I--40129,
   Bologna, Italy \\
   $^3$Dipartimento di Astronomia, Universit\`a di Bologna, Via
   Ranzani 1, I--40127, Bologna, Italy\\
   $^4$Laboratorio de Astrof\'isica Espacial y F\'isica Fundamental
   (CAB-CSIC-INTA), Postal Address: \\
   LAEFF, European Space Astronomy Center,
   P.O. Box 78, E-28691 Villanueva de la Ca\~nada, Madrid \\
   $^5$Institute of Astronomy, Madingley Road, Cambridge CB3 0HA\\
   $^6$SRON Netherlands Institute for Space Research, Sorbonnelaan 2, 
   3584 CA Utrecht, The Netherlands\\
   $^7$Astronomical Institute, University of Utrecht, Postbus 80000, 
   3508 TA Utrecht, The Netherlands \\ 
   $^8$Laboratoire d'Astrophysique de GrenobleÐ-Universit\'e 
   Joseph--Fourier/CNRS UMR 5571 Ð-BP 53, F--38041 Grenoble, France \\
   $^9$Dipartimento di Fisica, Universit\'a degli Studi Roma Tre, 
   via della Vasca Navale 84, 00146 Roma, Italy \\
   $^{10}$INAF/Osservatorio Astronomico di Brera, Via Brera 28, 20121, Milano, Italy \\
}

\pagerange{\pageref{firstpage}--\pageref{lastpage}}

\usepackage{times}

\begin{document}

\label{firstpage}

 \maketitle

\begin{abstract}
We report on partially overlapping \xmm\ ($\sim$260 ks) and
\suzaku\ ($\sim$100 ks) observations of the iron K band in the nearby,
bright Seyfert 1 galaxy Mrk~509.  The source shows a resolved neutral
Fe K line, most probably produced in the outer part of the accretion
disc.
Moreover, the source shows further emission blue--ward of the 6.4 keV
line due to ionized material. This emission is well reproduced by a
broad line produced in the accretion disc, while it cannot be easily
described by scattering or emission from photo--ionized gas at
rest. The summed spectrum of all \xmm\ observations shows the presence
of a narrow absorption line at 7.3 keV produced by highly ionized
outflowing material. A spectral variability study of the \xmm\ data shows an
indication for an excess of variability at 6.6--6.7 keV. 
These variations may be produced in the red wing of the broad ionized 
line or by variation of a further
absorption structure. The \suzaku\ data indicate that the neutral 
Fe K$\alpha$ line intensity is consistent with being constant on long 
timescales (of a few years) and they also confirm as most likely the 
interpretation of the excess blueshifted emission in terms of a broad 
ionized Fe line. The average \suzaku\ spectrum differs from the \xmm\ 
one for the disappearance of the 7.3 keV absorption 
line and around 6.7 keV, where the \xmm\ data alone suggested
variability. 
\end{abstract}


\section{Introduction}

Deep investigations of the Fe K band in the brightest AGNs allow us to
probe the presence of highly ionized emitting/absorbing components
from the innermost regions around the central black hole.
The high--sensitivity X--ray satellites \xmm\ and
\chandra\ have shown that the presence of a narrow core of the lowly ionized
Fe K$\alpha$ line is nearly ubiquitous (Yaqoob \& Padhmanaban 2004;
Guainazzi et al. 2006; Nandra et al. 2007) and that ionized components
of the line, generally associated with emission from photo-- and/or
collisionally-- ionized distant gas are also common (NGC 5506, 
NGC 7213, IC 4329A; Bianchi et al. 2003; Page et al. 2003; 
Reynolds et al. 2004; Ashton et al. 2004; Longinotti et al. 2007; 
see also Bianchi et al. 2002; 2005). 
The presence of broad (neutral or ionized) components
of Fe K lines can only be tested via relatively long exposures of the
brightest sources (e.g., Guainazzi et al. 2006; Nandra et al. 2007).  Moreover, the
observational evidence for broad lines and their interpretation in
terms of relativistic effects may be questioned when an important
absorbing ionized component is present.  Spectral variability studies help in
disentangling the different, often degenerate, spectral components (Ponti et
al. 2004; Iwasawa et al. 2004; Ponti et al. 2006; Tombesi et al. 2007;
Petrucci et al. 2007; DeMarco et al., in prep.).

Mrk~509 (z=0.034397) is the brightest Seyfert 1 of the hard (2--100 keV) X-ray sky
(Malizia et al. 1999; Revnivtsev et al. 2004; Sazonov et al. 2007) 
that is not strongly affected by a warm absorber
component (Pounds et al. 2001; Yaqoob et al. 2003).  The HETG \chandra\
observation confirms the presence of a narrow component of the Fe K
line with an equivalent width (EW) of 50 eV (Yaqoob et al. 2004).  The
presence of a second ionized component of the Fe K line at 6.7--6.9 keV
has been claimed by Pounds et al. (2001) who fitted it using a relativistic 
profile, but Page et al. (2003) showed that the same spectral feature 
was consistent also with a simple Compton reflection component from 
distant material. The broad--band \sax\ spectrum and, in particular, the soft
excess, have been fitted by De Rosa et al. (2004) with a reflection
component from a ionized disc in addition to a neutral reflection
component.  Finally, Dadina et al. (2005) found evidence of absorption
due to transient, relativistically red--blue-- shifted ionized matter.

Here we present the spectral and variability analysis 
of the complex Fe K band of Mrk~509, using the whole set of \xmm\ and 
\suzaku\ observations. 
The paper is organized as follows. Section 
2 describes the observations and the data reduction. 
In Sect. 3 the spectral analysis of the EPIC-pn data of the Fe K band 
(using phenomenological models) is presented. In particular 
in Sect. 3.3, to check
for the presence of an absorption line, the EPIC-MOS data have 
also been considered. In Sect. 3.4 the spectral variability 
analysis, within the \xmm\ observations, is presented. Sect. 4 describes the 
spectral analysis of the Fe K band of the \suzaku\ summed (XIS0+XIS3) data 
and the detailed comparison with the spectrum accumulated during 
the \xmm\ observations. In Sect. 4.1 the HXD-pin data are 
introduced in order to estimate the amount of reflection continuum 
present in the source spectrum. 
Finally, a more physically self-consistent fit of the EPIC spectra 
of all the EPIC instruments (EPIC-pn plus the two EPIC-MOS)
is investigated in Sect. 5. 
The results of our analysis are discussed in Sect. 6, followed by 
conclusions in Sect. 7.

\section{Observations and data reduction}

Mrk~509 was observed 5 times by \xmm\ on 2000--10--25, 2001--04--20,
2005--10--16, 2005--10--20 and 2006--04--25.
All observations were performed with the EPIC--pn CCD camera operating
in small window observing mode and with the thin filter applied. 
The total pn observation time is of about 260 ks.
Since the live--time of the pn CCD in small window mode is 71 per cent,
the net exposure of the summed spectrum is of about 180 ks. 
The analysis has been made with the SAS software (version 7.1.0), 
starting from the ODF files. Single
and double events are selected for the pn data, while only single
events are used for the MOS camera because of a slight pile--up effect.
For the pn data we checked that the results obtained using only single
events (that allow a superior energy resolution) are consistent with 
those from the MOS, finding good agreement. The source and background 
photons are extracted from a region of 40 arcsec within the same CCD 
of the source both for the pn and MOS data. Response matrices were 
generated using the SAS tasks {\sc RMFGEN} and {\sc ARFGEN}.

\suzaku\ observed Mrk~509 four times on 2006--04--25, 2006--10--14,
2006--11--15 and 2006--11--27. The last \xmm\ and the first
\suzaku\ observations overlap over a period of $\sim$~25~ks.  
Event files from version 2.0.6.13 of the \suzaku\ pipeline processing 
were used and spectra were extracted using {\sc XSELECT}.
Response matrices and ancillary response files were generated for 
each XIS using {\sc XISRMFGEN} and {\sc XISSIMARFGEN} version 
2007--05--14.
The XIS1 camera data are not considered here because of the 
relatively low effective area in the Fe K energy interval, 
while the XIS2 is unavailable for observations performed after 
November 2006.
We used the data obtained during the overlapping interval to check whether the 
EPIC pn and MOS data on one hand and the \suzaku\ XIS0 and XIS3 data on the other hand 
are consistent within the inter--calibration uncertainties. 
We found an overall good agreement between the data from the two satellites, 
the parameters related to the main iron emission features and the power--law 
continuum being the same within the errors (except for the XIS2 camera above 8 keV).
The total XIS observation time is about 108 ks.
The source and background photons are extracted from 
a region of 4.3 arcmin within the same CCD of the source.
For the HXD/PIN, instrumental background spectra and response 
matrices provided by the HXD instrument team have been used. 
An additional component accounting for the CXB has been included
in the spectral fits of the PIN data.

All spectral fits were performed using the Xspec software
(version 12.3.0) and include neutral Galactic absorption 
(4.2$\times$10$^{20}$~cm$^{-2}$; Dickey \& Lockman 1990), the energies
are rest frame if not specified otherwise, and the errors are reported
at the 90 per cent confidence level for one interesting parameter (Avni
1976).
The sum of the spectra has been performed with the {\sc 
MATHPHA}, {\sc ADDRMF} and {\sc ADDARF} tools within the 
{\sc HEASOFT} package (version 6.1).

\section{Fe K band emission of Mrk~509: the \xmm\ data}

The primary goal of this investigation is the study of the Fe K line
band; therefore, in order to avoid the effects of the warm absorber
(although not strong; Yaqoob et al. 2003; Smith et al. 2007) and of
the soft excess, we concentrate on the analysis of the data in the
3.5--10 keV band only. A detailed study of the warm absorber and 
its variations will be performed by Detmers et al. (in prep), 
we can nevertheless anticipate that the warm absorber has negligible effect 
in the Fe K energy band and thus on the results presented here.
\begin{figure}
\includegraphics[width=0.28\textwidth,height=0.46\textwidth,angle=-90]{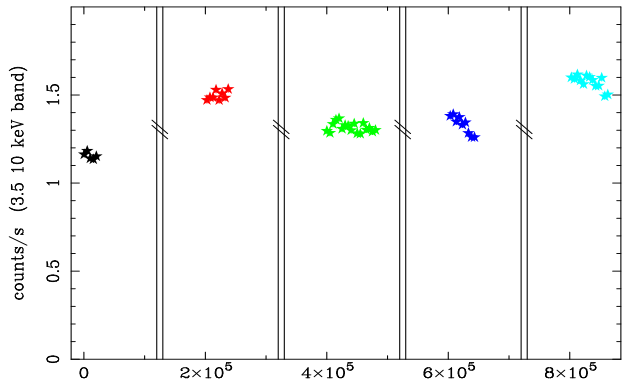}
 \caption{3.5--10 keV EPIC--pn light curves of the \xmm\ observations. The abscissa shows the 
observation time in seconds. The time between the different observations is arbitrary. 
The black, red, green, blue and light blue show the light curves during the 2000--10--25, 
2001--04--20, 2005--10--16, 2005--10--20 and 2006--04--25 observations, respectively.} 
\label{lightcurve}
\end{figure}
Figure \ref{lightcurve} shows the source light curve in the 3.5--10 keV
energy band obtained from the \xmm\ pointings. 
Mrk~509 shows
variations of the order of $\sim$30 per cent over the different observations,
while almost no variability is detected within each observation.
Only during the fourth observation the source shows significant variability, 
with a mean fractional rms of about 0.04.

\begin{figure}
\includegraphics[width=0.2\textwidth,height=0.46\textwidth,angle=-90]{pnlrat.ps} 
\vspace*{-0.05cm}

\includegraphics[width=0.46\textwidth,height=0.34\textwidth]{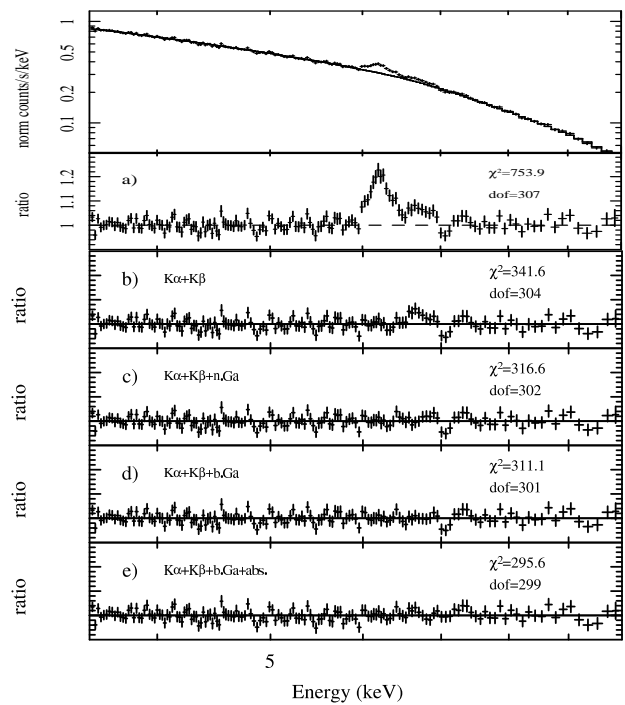}
 \caption{{\it (Upper panel)} Observed--frame 3.5--10 keV summed \xmm\ 
   EPIC--pn spectrum fitted in the 3.5--5 and 8--10 keV band with a power law. 
   {\it (Panel a)} Data/model ratio.  This ratio
   shows a clear evidence for a neutral Fe K emission line and further
   emission from ionized Fe, as well as other complexities around 7
   keV. {\it (Panel b)} Data/model ratio when two resolved emission
   lines (for the Fe K$\alpha$ and K$\beta$) are included in the
   spectral fitting.  Strong residuals are still present, indicative
   of ionized Fe K emission, while no residual emission redward of the
   neutral Fe K line appears. {\it (Panel c)} Data/model ratio when a
   narrow emission line is included in the model to reproduce the
   ionized emission. {\it (Panel d)} Same as panel c, but with a
   single broad emission line instead of a narrow line. In both
   cases (panel c and d), an absorption feature is present
   around 7 keV. {\it (Panel e)} Data/model ratio when an absorption
   component (modeled using {\sc xstar}) and a relativistic ionized
   line are added to the power law and the emission from neutral Fe
   K.}
\label{res_FeK}
\end{figure}

We start the analysis of the \xmm\ data considering the spectra from 
the EPIC-pn camera only (including the EPIC-MOS 
data only when a check of the significance of a feature is required; see Sect. 3.3).
We have fitted a simple power law model to the 3.5--10 keV data and found that the 
spectral index steepens with increasing flux. 
It goes from 1.54$\pm$0.03 to 1.72$\pm$0.03 for fluxes of
2.5$\times$10$^{-11}$ and 3.3$\times$10$^{-11}$ erg
cm$^{-2}$ s$^{-1}$, respectively (3.0$\times$10$^{-11}$ --
4.3$\times$10$^{-11}$ erg cm$^{-2}$ s$^{-1}$, in the 2--10 keV 
band). 
We firstly phenomenologically fitted the Fe K complex of each single 
observation with a series of emission-absorption lines (see also $\S$3.3) 
and checked that the results on the parameters of Fe K complex obtained 
in each observation are consistent within the errors (not a surprising result 
in light of the low statistics of the single spectra and weakness of 
the ionized features; see $\S$ 3.4). 
Hence, we concluded that the continuum variations do not 
strongly affect the observed shape of the narrow--band emission/absorption 
structures in the Fe K band. 
Thus, in order to improve the signal--to--noise ratio and thus 
to detail the fine structures of the Fe K band, the spectra 
of all the \xmm\ observations have been summed 
(see $\S$\ref{spec_var} for the study of the 
source spectral variability).  The summed mean EPIC--pn 
spectrum has been grouped in order to 
have at least 1000 counts in each data bin. Moreover, this binning
criterion ensures to have at least 30 data--points per keV in the 4--7
keV band, where the Fe K$\alpha$ complex is expected to
contribute. This guarantees a good sampling of the energy
resolution of the instrument and the possibility of fully exploiting
the spectral potentials of the EPIC instruments.  Fig. \ref{res_FeK}
shows the ratio between the data and the best--fit power law.  The
energy band used during the fit has been restricted to 3.5--5 and 8--10
keV, in order to avoid the Fe K band, hence measuring the underlying
continuum.
The resulting best--fit power--law continuum has a photon index of
1.63$\pm$0.01 and very well reproduces the source emission 
($\chi^2$=170.0 for 163 degrees of freedom, dof) outside the 
Fe K band. The inclusion of the Fe K band  
shows that other components are necessary to reproduce it 
($\chi^2$=753.9 for 307 dof). The bad
statistical result is explained by the presence of clear spectral
complexity in the 6--7~keV band. 

\subsection{The 6.4 keV emission line}

\begin{figure}
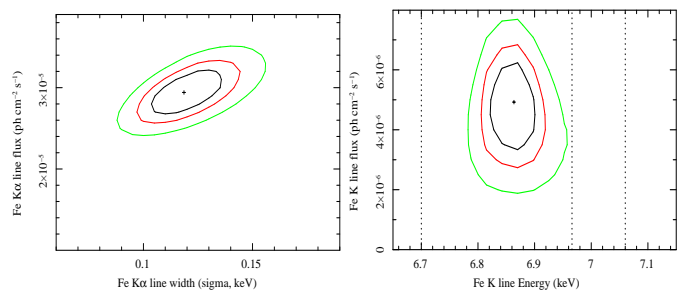

\hspace{-0.4cm}
\includegraphics[width=0.23\textwidth,height=0.25\textwidth,angle=-90]{3lines_sigma64_2.ps} 
\includegraphics[width=0.23\textwidth,height=0.25\textwidth,angle=-90]{FeK_2_2.ps} 
 \caption{{\it (Left panel)} Contour plot of the sigma vs. intensity
   of the neutral Fe K line.  {\it (Right panel)} Contour plot of the
   energy vs. intensity of the narrow line used to fit the ionized Fe
   K emission. The narrow line energy is not consistent with emission
   from Fe XXV (neither with the forbidden at 6.64 keV, nor with the 
   resonant at 6.7 keV), Fe XXVI or Fe K$\beta$, whose energy is indicated by
   the vertical dotted lines (from left to right).}
\label{cont3lines}
\label{sigma64}
\end{figure}

Panel a of Fig. \ref{res_FeK} shows the clear evidence for a prominent emission 
line, consistent with a neutral Fe K$\alpha$ line at 6.4 keV. We therefore
added a Gaussian emission line to the model, obtaining a very
significant improvement of the fit ($\Delta\chi^2$=392.1 for
the addition of 3 dof). The best--fit energy of the line is
6.42$\pm$0.02 keV, consistent with emission from neutral or slightly 
ionised material. The line has an equivalent width of 69$\pm$8 eV 
and is clearly resolved ($\sigma$=0.12$\pm$0.02 keV), as shown
by the contour plot in the left panel of Fig. \ref{sigma64}.
The residuals in panel b of Fig. \ref{res_FeK} show no excess
redward of this emission line, which could have been indicative of
emission from relativistically redshifted neutral material. 

\subsection{The ionized Fe K emission line}

An excess is, however, present in the range 6.5--7 keV (Fig.
\ref{res_FeK}, panel a). If modeled with a Fe K$\beta$ component with
the expected energy (fixed at 7.06 keV) and forced to have an
intensity of 0.15 of the K$\alpha$ (Palmeri et al. 2003a,b; Basko 1978; 
Molendi et al. 2003) and a width equal to the Fe K$\alpha$ line 
(i.e. assuming that the K$\alpha$ and K$\beta$ line originate from 
one and the same material), the fit improves significantly 
($\Delta$$\chi^2$=20.3). Nonetheless, significant residuals are 
still present in the 6.5--6.9 keV band (panel b of Fig.\ref{res_FeK}). 
If this further excess is modelled with a narrow
Gaussian line ($\Delta\chi^2$=25 for 2 additional dof), the feature
(EW=12$\pm$4 eV) is found to peak at E=6.86$\pm$0.04 keV (see panel c of Fig. 2 
and right panel of Fig. \ref{cont3lines}).  Thus, the energy centroid is not
consistent with the line being produced by either Fe XXV or Fe XXVI
(right panel of Fig. \ref{cont3lines}) in a scattering medium distant
from the X-rays source (Bianchi et al. 2002; 2004). 
The higher energy transition of the Fe XXV complex is the "resonant line" expected 
at 6.7 keV (see e.g. Bianchi et al. 2005). 
Thus, to save this interpretation, it is required that the photo-ionized gas
has a significant blueshift ($\sim$5700 km/s, if the line is associated to
Fe XXV) or redshift ($\sim$4500 km/s, for Fe XXVI).
Then, instead of fitting the ionized excess with a single line, we 
fitted it with two narrow lines forcing their energies 
to be 6.7 and 6.96 keV. The fit clearly worsens ($\chi^2$=326.7 for 302 dof, 
corresponding to a $\Delta$$\chi^2=-10.1$ for the same dof). 
However, if the gas is allowed to be outflowing, the fit improves 
($\Delta$$\chi^2$=4.3 for the addition of 1 new parameter; 
$\chi^2$=312.3 for 301 dof; the EW are 8.9 and 12.4 eV for the Fe XXV 
and Fe XXVI lines, respectively) as respect to the single narrow 
emission line and it results to have a common velocity of 
3500$^{+1900}_{-1200}$ km/s. 

Alternatively, the excess could be produced by a single broad line
coming from matter quite close to the source of high--energy photons (in this case 
the Fe K emission is composed by Fe K$\alpha$+$\beta$ plus another 
Fe K line). Leaving the width of the line free to vary, the fit improves, 
with $\chi^{2}$=311.1 (panel d Fig. 2) and $\Delta\chi^{2}$ of 5.5, with 
respect to the single narrow ionized emission line fit, and 
$\Delta\chi^{2}$ of 1.3 for the same dof with respect to the  
best--fit model with two narrow ionized lines.
The resulting broad ionized Fe K line has EW=23$\pm$9~eV
and $\sigma$=0.14$^{+0.13}_{-0.08}$ keV.  The best--fit energy of the
line does not change significantly (E=6.86$^{+0.08}_{-0.16}$
keV); however, in this case the emission is consistent (at the 99 per cent
confidence level) with either Fe XXV or Fe XXVI.  Although the
statistical improvement is not highly significant, in the following we
will consider that the $\sim$6.8--6.9 keV excess is indeed associated
with a resolved emission line.

\subsection{Ionized absorption?}

\begin{figure}
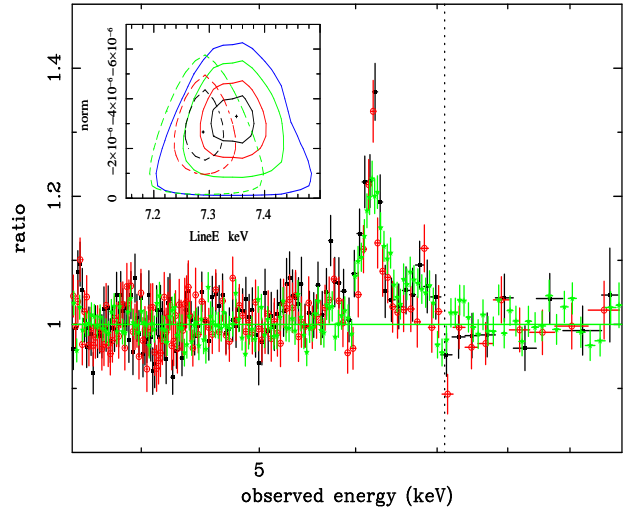

\includegraphics[width=0.38\textwidth,height=0.46\textwidth,angle=-90]{pnM1M2_pat0.ps} 
\hspace*{-7.35cm}
\vspace*{-3.1cm}
\includegraphics[width=0.18\textwidth,height=0.18\textwidth,angle=-90]{FeKalfabeta_abs_signafreezeMOSpn2_2.ps}
\vspace*{+3.3cm}
 \caption{Superposition of the pn (green), MOS1 (black) and MOS2 (red)
   summed spectra of all the \xmm\ observations. The data are fitted,
   in the 3.5--5 and 8--10 keV bands with a power law, absorbed by
   Galactic material.  The same structures are present in the three
   spectra.  In particular, a narrow drop of emission is present 
   in all the instruments at the same energy (see vertical
   dotted line).  {\it (Inset panel)} Confidence contour plot of the intensity
   vs. energy of the narrow unresolved ionized absorption when using
   the pn data alone (dashed contours) and including the
   MOS data as well (solid contours).  The lines indicate the 68.3 (black),
   90 (red), 99 (green) and 99.9 (blue) per cent confidence levels.}
\label{MOS}
\end{figure}

The \xmm\ data also display a narrow absorption feature at
E$\sim$7~keV (observed frame; see Fig. 2, panel d). Since this
feature is very close to the broad excess we just discussed, its
significance and intensity are degenerate with the broad emission--line
parameters. In order to gain some insight, we then fixed the broad 
emission--line parameters at the best--fit ones obtained before the addition of
a narrow ($\sigma$ fixed at 1 eV) Gaussian absorption line component. 
In this case, the line is significant at the $\sim$99 per cent confidence 
level (dashed contours of Fig. \ref{MOS}; $\Delta\chi^2$=15.5 for 2 additional 
parameters; see also panel e of Fig. 2). Once the MOS
data\footnote{ The shapes of the emission/absorption lines in the MOS
  instruments appear slightly narrower, although consistent with the
  values obtained with the pn instrument. } are added, the
significance of this feature increases to 99.9 per cent (solid contours of
Fig. \ref{MOS}), in both cases, of a broad and of a narrow ionized
emission line. The best fit energy and EW of the line are 
E=7.28$^{+0.03}_{-0.02}$ keV and EW=$-$14.9$^{+5.2}_{-5.5}$ eV,
E=7.33$^{+0.03}_{-0.04}$ keV and EW=$-$13.1$^{+5.9}_{-2.9}$ eV,
in the pn alone and in the pn+MOS, respectively.

\subsection{Time resolved spectral variability and total rms spectrum}
\label{spec_var}

One of the goals of the present analysis is to search for
time--variation of the emission/absorption features of the Fe K
complex.  To measure possible variations in the Fe K band, the mean
EPIC-pn spectra of each of the 5 \xmm\ observations have 
been studied.  The spectra are fitted with the same model 
composed by a power law plus three emission lines 
for the Fe K$\alpha$, K$\beta$ (with the width fixed at 
the best--fit value, $\sigma$=0.12 keV) and the broad ionized
Fe K line. The low statistics of the spectra of the single observations 
prevents us from the detection of significant spectral variability of 
the weak ionised emission/absorption lines. The neutral Fe K line 
is better constrained and we find that its EW is anti--correlated with the
level of the continuum, as expected for a  constant line. 

\begin{figure}
\hspace{-0.16cm}
\includegraphics[width=0.12\textwidth,height=0.46\textwidth,angle=-90]{ratio.ps}
\vspace*{-0.1cm}

\includegraphics[width=0.46\textwidth,height=0.3\textwidth]{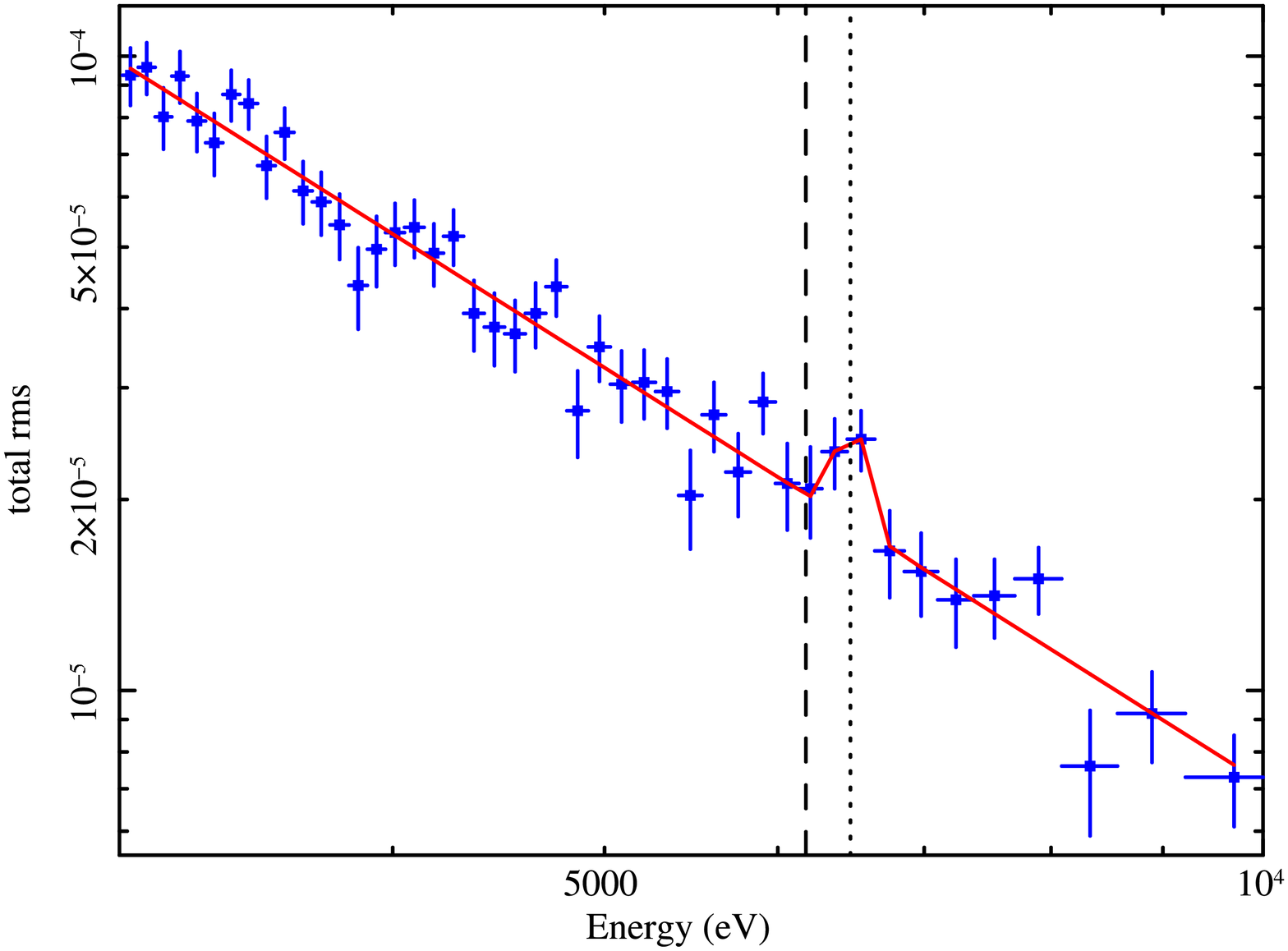}
 \caption{
   {\it Lower panel:} Total rms variability spectrum of the
   \xmm\ observations.  The data (blue crosses) show the spectrum of the variable
   component. The best--fit model is a power law with spectral index
   $\Gamma$=2.18 (red line) plus a Gaussian emission line
   (improving the fit by $\Delta\chi^2$ of 8.9 for the addition of 2
   parameters).  The dashed line highlights the centroid energy of the
   neutral Fe K$\alpha$ line, while the dotted line is placed at the
   maximum of the variability excess, modeled with the Gaussian
   emission line. The excess variability energy corresponds to a drop
   of emission of the real spectrum.}
\label{fvar}
\end{figure}

A different, more sensitive, way to detect an excess of spectral 
variability is the total rms function.
The upper panel of Fig. \ref{fvar} displays the shape of the
summed spectrum in the Fe K line band.  The lower panel shows the
total rms spectrum (Revnivtsev et al. 1999; Papadakis et al. 2005)
calculated with time bins of $\sim$4.5 ks.  
The total rms is defined by the formula:
\begin{equation}
RMS(E)=\frac{\sqrt{S^2(E)-<\sigma^2_{err}>}}{\Delta E * arf(E)}
\end{equation}
where S$^2$ is the source variance in a given energy interval $\Delta$E; 
$<\sigma^2_{err}>$ is the scatter introduced by the Poissonian 
noise and {\it arf} is the telescope effective area 
convolved with the response matrix\footnote{The total rms spectrum 
provides the intrinsic source spectrum of the variable component.
Nevertheless, we measure the variance as observed through the instrument.
Thus, the sharp features in the source spectrum, as well as the effects of 
the features on the effective area, are broadened 
by the instrumental spectral resolution. For this reason, to 
obtain the total rms spectrum, we take into account the convolution of the 
effective area with the spectral response.}.
This function shows the spectrum of the varying component only, in which any constant
component is removed and has been computed by using the different \xmm\ 
observations as if they were contiguous. 
The total rms spectrum may be reproduced by a power law with
a spectral index of 2.13 ($\chi^2$=46.4 for 43 dof).  Thus, the variable
component is steeper than the observed power law in the mean spectrum,
in agreement with the mentioned observed steepening of the photon
index ($\Gamma$) with flux. The $\Gamma$--flux correlation is commonly 
observed in Seyfert galaxies and has been interpreted as being due to the 
flux-correlated variations of the power-law slope produced in a 
corona above an accretion disc and related to the changes in the input 
soft seed photons (e.g. Haardt, Maraschi \& Ghisellini 1997; Maraschi 
\& Haardt 1997; Poutanen \& Fabian 1999; Zdziarski et al. 2003). 
These models predict the presence of a pivot point, that would 
correspond to a minimum in the total rms spectrum. The observation 
of a perfect power law shape (see Fig. 5) indicates that the pivot point 
(if present) has to be outside the 3--10 keV energy band.
On the other hand, the slopeÐ-flux behaviour can be explained 
in terms of a two-component model (McHardy, Papadakis \& Uttley 1998; 
Shih, Iwasawa \& Fabian 2002) in which a constant-slope power law 
varies in normalization only, while a harder component remains approximately 
constant, hardening the spectral slope at low flux levels only, when 
it becomes prominent in the hard band. In this scenario the spectral index 
of the variable component is equal to the one of the total 
rms spectrum, that is $\Gamma$=2.13. 

Moreover we note that at the energy of the neutral and ionized Fe K line 
components, no excess of variability is present, in agreement 
with these components being constant, while an indication for an 
excess of variability is present around 6.7 keV. 
In order to compute the significance 
of this variability feature, a narrow Gaussian line has been 
added to the modelling of the total rms spectrum. The best--fit 
energy of the additional line is 6.69 keV, with
a $\sigma$ fixed at the instrumental energy resolution, while the 
resulting $\Delta\chi^2$ is 8.9 for the addition of 2 parameters 
(that corresponds to an F--test significance of 98.8 per cent).
Introducing the line, the 
continuum spectral index steepens to $\Gamma \sim 2.18$. 
The dashed line in Fig. 5 highlights the centroid energy of the neutral Fe K$\alpha$
line, while the dotted line (at $\sim$6.7 keV, rest frame) is placed at the
maximum of the variability excess. This energy corresponds to a drop
of emission in the real spectrum, as we shall discuss in more detail
in Section 5. 

\section{The \suzaku\ view of the Fe K band emission}

As mentioned in $\S$2, the source was also observed with Suzaku. 
The first 25 ks \suzaku\ observation is simultaneous with the 
last \xmm\ pointing. The source spectra of all the instruments
are in very good agreement, during the simultaneous 
observation. The spectrum is also consistent with the presence 
of the emission and absorption lines, as observed in the mean 
\xmm\ spectrum, nevertheless, due to the low statistics 
of the 25 ks spectrum and the weakness of the ionized 
features, it is not possible to perform a detailed comparison.
Only the presence of the strong Fe K$\alpha$ line can be 
investigated, the ionized emission and absorption lines are 
not constrained in the 25 ks \suzaku\ exposure.

\begin{figure}
\includegraphics[width=0.3\textwidth,height=0.46\textwidth,angle=-90]{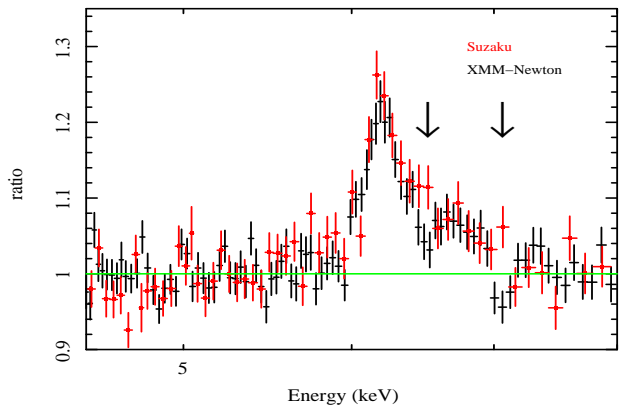}
 \caption{\xmm\ (black) and \suzaku\ XIS0+XIS3 (red) summed mean
   spectra. The data are fitted, in the 3.5--5 and 7.5--10 keV bands,
   with a simple power law, absorbed by Galactic material, and the ratio of the 
   data to the best fit model is shown. The arrows mark absorption features 
   in the spectrum. 
}
\label{SuzakuXMM}
\end{figure}

Also during the 4 \suzaku\ pointings, Mrk~509 has shown little 
variability, with flux changes lower than 10--15 per cent, hampering 
any spectral variability study.
Fig. \ref{SuzakuXMM} shows the \xmm\ (black) and \suzaku\ XIS0+XIS3
(red) summed mean spectra.  The data were fitted, in the 3.5--5 and
7.5--10 keV bands, with a simple power law and Galactic absorption: 
the ratio of the data to the best fit model is shown in Fig. 6. 
The source emission varied between the \xmm\ and the \suzaku\ observations.
The best--fit spectral index and the 3.5--10 keV band fluxes are:
$\Gamma$=1.63$\pm$0.01 and $\Gamma$=1.71$\pm$0.02 and
2.63$\times$10$^{-11}$ and 3.11$\times$10$^{-11}$ ergs cm$^{-2}$
s$^{-1}$, during the \xmm\ and \suzaku\ observations, respectively. 
The neutral and ionized Fe K emission lines appear constant, while
some differences are present at 6.7 keV, the same energy where the
\xmm\ data were suggesting an increase of variability.  Other more
subtle differences appears at $\sim$~7 keV, where the absorption line
imprints its presence in the \xmm\ data only.

The \suzaku\ spectrum of Mrk~509 shows, in good agreement with the
\xmm\ one, a resolved neutral Fe K line smoothly joining with a higher
energy excess, most likely due to ionized iron emission (see Fig.
\ref{SuzakuXMM}).  Given that no absorption lines around 6.7 keV or
7.3 keV are present in the \suzaku\ data, the spectrum may be useful
to infer the properties of the emission lines more clearly. 

The XIS0+XIS3 \suzaku\ summed spectrum has been fitted in the 3.5--10
keV band with a power law plus two resolved Gaussian emission lines 
to reproduce the emission from Fe K$\alpha$+$\beta$. 
The parameters of the Fe K$\alpha$ line are free to vary, while the 
Fe K$\beta$ ones are constrained as in $\S$3.2.
This fit leaves large residuals ($\chi^2$=1379.8 for 1337 dof) in the 
Fe K band. In this respect, it is difficult 
to describe the $>$6.5 keV excess with a single narrow ionized 
Fe line (either due to Fe XXV or Fe XXVI). In fact, although the addition of a 
narrow line is significant ($\Delta\chi^2$=20.1 for 
2 more parameters), it leaves residuals in the Fe K band. This remaining excess 
can be reproduced ($\Delta\chi^2$=5.9 for 1 more parameter), in a photoionized 
gas scenario, by a blend of two unresolved ionized lines, requiring three emission 
lines to fit the Fe K band (FeK$\alpha$+$\beta$, Fe XXV and Fe XXVI). 
In this case, such as in the analysis of the \xmm\ mean spectrum, 
a blueshift of this component is suggested 
(v=2600$^{+2800}_{-2000}$ km s$^{-1}$).
However, the best--fit model (this scenario is strengthened by the lack of 
narrow peaks) suggests that the excess may be in fact associated with a broad ionized Fe
line (over which the $\sim$~6.7 keV and $\sim$~7.3 keV absorption lines
are most likely superimposed, but during the \xmm\ observation only).
In fact considering a broad Fe line instead of the two narrow lines we obtain 
an improvement of $\Delta\chi^2$=9.7 for the same dof  (see Table 
\ref{Suzaku}, model A).
\begin{table*}
\small
\begin{tabular}{|cc|cccc|cccc|c|cccc}
\hline 
\hline 
&3.5--10 keV & BEST--FIT & SPECTRA\\
\hline 
\hline 
& Suzaku\\
& $\Gamma$&pl norm$^{\rm a}$ &E$_{\rm Neut.}$&$\sigma$$_{\rm Neut.}$&A$_{\rm Neut.}$$^{\rm b}$ (EW)$^{\rm c}$& E$_{\rm Ion.}$ & $\sigma$$_{\rm Ion.}$/r$_{\rm in}$ & A$_{\rm Ion.}$$^{\rm b}$ (EW)$^{\rm c}$ & $\chi^2$/dof \\
& &&keV&keV& & keV & keV/r$_{\rm g}$& \\
A& 1.72$\pm$0.02&1.12$\pm$0.02&6.42$\pm$0.03&$<$0.06&1.7$\pm$0.5 (32)&6.54$\pm$0.09&0.40$\pm$0.1&4.6$\pm$1.2 (90)&1344/1340\\
B& 1.72$\pm$0.02&1.12$\pm$0.02&6.42$\pm$0.02&$<$0.07&2.1$\pm$0.5 (40)&6.61$\pm$0.08&24$\pm$10&3.4$\pm$0.8 (79)&1346/1340\\
\hline
&Self-consistent model \\
&XMM-Newton\\
&$\Gamma$&pl norm$^{\rm a}$&E$_{\rm Neut.}$&$\sigma$$_{\rm Neut.}$&A$_{\rm Neut.}$$^{\rm b}$
 &Incl. & $\xi$ & A$_{\rm Refl.Ion.}$$^{\rm b}$\\
&&&keV&keV& & deg & erg cm s$^{-1}$ & \\
C& 1.70$\pm$0.01&0.92$\pm$0.04&6.41$\pm$0.01&0.07$\pm$0.01&2.2$\pm$0.3 &47$\pm$2&11$^{+200}_{-7}$&0.9$^{+3.0}_{-0.5}$ \\
& N$_H$$^{\rm d}$ & log($\xi$) & z & $\chi^2$/dof \\ 
& 5.8$^{+5.2}_{-4.8}$&5.15$^{+1.25}_{-0.52}$&$-$0.0484$^{+0.012}_{-0.013}$&894.3/876\\
\hline
\end{tabular}
\caption{{\it Top panel:} Best--fit values of the summed spectra (XIS0+XIS3) of all \suzaku\ 
observations fitted in the 3.5--10 keV band. Both model A and B include a power law and two 
Gaussian lines K$\alpha$+$\beta$ to fit the 6.4 keV excess. In addition to this baseline model, either 
another Gaussian component (Model A) or a {\sc DISKLINE} profile (Model B) have been added 
to reproduce the ionized line, respectively. 
In Table the best--fit power law spectral index ($\Gamma$) and normalization as well as the 
Fe K$\alpha$ energy, width and normalization are reported for model A and B. The 
energy, width and normalization are reported when a Gaussian profile for the ionized 
Fe K line is considered (Model A), while the best fit energy, inner radius and normalization are 
presented when a {\sc DISKLINE} profile is fitted (Model B). Standard disc reflectivity index, 
outer disc radius and disc inclination of $\alpha=3$, r$_{out}=400$ r$_g$ and 30$^{\circ}$
have been assumed for the relativistic profile.
{\it Bottom panel:} Best fit results of the summed \xmm\ EPIC-pn and EPIC-MOS data of Mrk~509 
(fitted in the 3.5--10 keV band). The model 
({\sc wabs*zxipcf*(pow+zgaus+zgaus+pexrav+kdblur*(reflion))}) consists of: 
i) a power law; ii) two Gaussian 
emission lines for the Fe K$\alpha$ and K$\beta$ emission (this latter has energy is fixed 
to the expected value, 7.06 keV, intensity and width tied to the K$\alpha$ values); 
iii) a neutral reflection continuum component ({\sc pexrav} in Xspec) with $R$=1 (value 
broadly consistent with the pin constraints and the values previously observed; 
De Rosa et al. 2004), Solar abundance and high energy cut off of the illuminating 
power law at 100 keV; iv) a ionized disc reflection spectrum ({\it reflion} model; 
Ross \& Fabian 2005) with the disc inner and outer radii and the emissivity of
6, 400 r$_g$ and $-3$, respectively. The best fit disc inclination and ionization 
and the normalization of the disc reflection component are shown; v) an ionized 
absorption component ({\sc zxipcf}) totally covering the nuclear source. 
The best fit column density, ionization parameter and outflow velocity are 
reported.
{\it a) In units of 10$^{-2}$ photons keV$^{-1}$ cm$^{-2}$ s$^{-1}$ at 1 keV;
b) In units of 10$^{-5}$ photons cm$^{-2}$ s$^{-1}$;
c) In units of eV;
d) In units of 10$^{22}$ atoms cm$^{-2}$.}}
\label{Suzaku}
\end{table*} 

Thus, the \suzaku\ data indicate that the broad
excess at 6.5--6.6 keV is indeed due to a broad line rather than a
blend of narrow ionized Fe lines. Since broad lines may arise because
of relativistic effects in the inner regions of the accretion flow, we
tested this hypothesis by fitting the excess at 6.5--6.6 keV with a {\sc diskline}
profile.  The statistics of the spectrum is not such to allow us to
constrain all the parameters of the ionized {\sc diskline} model. Thus, the
disc reflectivity index has been fixed at the standard value ({\it $\alpha=-3$}, 
where the emissivity is proportional to $r^{\alpha}$), 
the outer disc radius and inclinations to 400 gravitational radii (r$_g$)
and 30$^{\circ}$, respectively.  The broad line is consistent with
being produced in the accretion disc (Table 1, Model B); however, the
emission from the innermost part of the disc is not required, the
lower limit on the inner disc radius being 10--15 r$_g$.  As clear 
from Fig.~\ref{SuzakuXMM}, the \suzaku\ data do not
require any ionized Fe K absorption structures.

In order to quantify the differences between the \suzaku\ and
\xmm\ spectra (and, in particular, the reality of the absorption structures at 6.7 and
7.3 keV appearing in the \xmm\ spectrum only) we fixed all the
parameters of the \suzaku\ model (apart from the intensity and
spectral index of the direct power law) and fit the \xmm\ data 
with that model. This corresponds to assuming that the intrinsic line
shapes do not vary between the two observations.  Then, a narrow
Gaussian line has been added to the \xmm\ model to estimate the
significance of the putative absorption structures.  The improvement in the spectral
fitting is evident, as indicated by the $\Delta\chi^2$=28.3 and 22 in
the case of a line at E=6.72$\pm$0.04~keV and E=7.29$\pm$0.04~keV,
respectively.  The presence of these spectral features only in the
\xmm\ observations is thus indicative of variability at energies
$\sim$6.6--6.7 and $\sim$7.3 keV.

\subsection{The \suzaku\ pin data to constrain the reflection fraction}

We add the pin data to measure the amount of reflection continuum. 
We note that the pin data provide a good quality spectrum up to 50 keV.
The model used involves a direct power law plus a neutral reflection 
component ($pexrav$ model in $Xspec$; Magdziarz \& Zdziarski 1995) 
plus the Fe K$\alpha$+$\beta$ resolved lines and a broad (DISKLINE) 
component of the line. 
As for model B we fix some of the parameters of the DISKLINE profile 
(disc inclination=30$^{\circ}$, r$_{out}$=400 r$_g$ and $\alpha$=-3).
Moreover we assume a high--energy cut off of 100 keV and Solar abundance. 
Thus, by fitting the 3--50 keV band data, we obtained a reflection 
fraction $R=0.4^{+0.6}_{-0.2}$ and a spectral index 
$\Gamma$=1.76$^{+0.12}_{-0.03}$. 
The total EW of the emission lines above the reflected continuum 
(about 1.2 keV) is broadly consistent with the theoretical 
expectations (Matt et al. 1996) and with what observed in Compton thick 
Seyfert 2 galaxies, where the primary continuum is absorbed and only 
the reflection is observed.
Nevertheless, also for this source, as already known from previous 
studies (Zdziarski et al. 1999), we observe that the spectral index 
and the reflection fraction are degenerate and strongly depend on 
the energy band considered. 
In fact, if the 2--10 keV band is considered, the reflection 
fraction increases, resulting to be $R$=1.1$^{+0.2}_{-0.5}$ and 
the power--law photon index of $\Gamma$=1.88$^{+0.03}_{-0.02}$. 
The total EW of the Fe emission lines above the reflected continuum 
are about 750 eV. Again these values are broadly in agreement with 
expectations (Matt et al. 1996). 

\section{A physically self-consistent fit: 
Possible origin of the spectral features}

The analysis of the \xmm\ and \suzaku\ data shows evidence for the
presence of: i) a resolved, although not very broad, ($\sigma\sim$0.12 keV)
neutral Fe K$\alpha$ line and associated Fe K$\beta$
emission; ii) an ionized Fe K emission line inconsistent with emission from
a distant scattering material at rest and most likely produced in the
accretion disc; iii) an absorption line at $\sim$7.3 keV, 
present in the summed spectrum of all \xmm\ observations only; 
iv) an indication for an enhancement of variability~-~both by
considering the \xmm\ data alone and by comparison between the two data
sets~-~at $\sim$6.7 keV that could be either due to the high
variability of the red wing of the broad ionized Fe K line, 
possibly associated with a variation of the ionisation of the disc,
or to a second ionized absorption line.

These emission/absorption components are partially inter--connected to each
other given the limited CCD resolution onboard \xmm\ and \suzaku. 
Thus we re--fit the \xmm\ (both the pn and MOS in the 3.5--10 keV 
energy band) data with a model containing components that better 
describe the physical processes occurring in the AGN. 
In particular, we consider two Gaussian lines for the Fe K$\alpha$ and 
K$\beta$ emission plus a neutral reflection component ({\sc pexrav} 
in {\sc xspec}) with a reflection fraction $R=1$ (consistent with the 
constraints given by the \suzaku\ pin data). The 
Fe K$\alpha$ line has an equivalent width of 1 keV above 
the reflection continuum.  
Moreover, we fit the broad ionized Fe K line with a fully 
self--consistent relativistic ionised disc reflection 
component ({\it reflion} model in {\it Xspec}; Ross \& Fabian 2005,
convolved with a {\sc LAOR} kernel; {\sc KDBLUR} in Xspec). 

The statistics prevents us from constraining the
parameters of the relativistic profile. 
Standard values for the relativistic profile are assumed, 
with the disc inner and outer radii and the emissivity of 6, 400 r$_g$, 
and $-$3, respectively. 
Finally, the $\sim$7.3 keV absorption line has been fitted with a photoionised
absorption model ({\it zxipcf} model in {\it Xspec}; Miller et
al. 2007; Reeves et al. 2008; Model C, Table 1), assuming a total 
covering factor.

Table \ref{Suzaku} shows the best--fit parameters. 
Once the presence of the reflection continuum is taken 
into account, the power law slope becomes steeper 
($\Gamma$=1.70$\pm$0.01, $\Delta\Gamma\sim$0.07) 
as compared to the fit with a simple power law and emission 
absorption lines (see \S 3).
The best fit energy of the neutral Fe K$\alpha$ line is 
E=6.41$\pm$0.01 keV, consistent with being produced by neutral material, 
and results to be narrower ($\sigma$=0.07$\pm$0.01 keV) 
than in the previous fits.
The ionized emission line is fitted with a ionized disc reflection model. 
The only free parameters of such a component are the inclination and 
ionisation parameter of the disc that result to be 47$\pm$2$^{\circ}$ and
$\xi$=11$^{+200}_{-7}$ erg cm s$^{-1}$ (Model C, Table 1). 
The material producing the 7.3 keV 
absorption feature in the \xmm\ data has to be highly ionized, 
as also indicated by the absence of a strong continuum curvature. 
In fact, the best ionization parameter is log($\xi$)=5.15$^{+1.25}_{-0.52}$ 
and the column density $N_H=5.8^{+5.2}_{-4.8}\times10^{22}$ cm$^{-2}$. 
Nevertheless the observed energy of the absorption feature does 
not correspond to any strong absorption features, thus there is 
evidence for this absorption component to be outflowing 
with a shift $v=-0.0484^{+0.012}_{-0.013}$ c 
($\sim14000^{+3600}_{-4200}$ km s$^{-1}$). 
The resulting $\chi^2$ is 894.3 for 876 dof. 

\section{Discussion}

This study clearly shows that long exposures are need to disentangle 
the different emitting/absorbing components contributing to the 
shape--variability of the Fe K complex in Seyfert galaxies. 
Here we discuss the origin of both neutral and 
ionized emission and absorption Fe lines in Mrk~509 which allow to have 
insights in the innermost regions of the accretion flow.

\subsection{Neutral/lowly ionized Fe emission line}

Once the broad ionized line is fitted, the width of the Fe K$\alpha$
line lowers to a value of 72$\pm$11 eV (see Fig. \ref{sigma64})
that corresponds to a FWHM(Fe K$\alpha$)=8000$\pm$1300 km s$^{-1}$
(see Model C, Table 1).
This value is slightly higher than that measured by Yaqoob \& Padmanabhan 
with a $\sim$50 ks HETG \chandra\ observation (2820$^{+2680}_{-2800}$ km s$^{-1}$).
The FWHM of the Fe K$\alpha$ line is larger than the width of the
H$\beta$ line (FWHM(H$\beta$)=3430$\pm$240 km s$^{-1}$; Peterson et
al. 2004; Marziani et al. 2003), indicating that the Fe line is
produced closer to the center than the optical BLR and, of course,
than the torus postulated in unified models; we note that a wide range 
of FWHM values is observed for the BLR and the Fe K lines in local Seyfert 
galaxies (Nandra 2006). 
However, the UV and soft X-ray spectra of Mrk 509 show evidence for 
the presence of broad emission lines with FWHM of
11000 km s$^{-1}$ (Kriss et al. 2000). The origin of these UV and
soft--X lines is still highly debated, nevertheless they may indicate
that the BLR region is stratified, i.e. that these lines are not 
produced in the optical BLR but in the inner part of a stratified
BLR region (see also Kaastra et al. 2002; Costantini et al. 2007), 
possibly as close as 2000 r$_g$ from the center (about
0.012 pc, being the mass of the black hole in Mrk 509
M$_{BH}\simeq$1.43$\pm$0.12$\times$10$^8$ M$_\odot$ Peterson et
al. 2004; Marziani et al. 2003). 
Nevertheless, if the line is produced in the innermost part of a stratified 
BLR, it would require either a higher covering fraction or a higher 
column density than generally derived from the optical 
and ultraviolet bands. Simulations by Leahy \& Creighton (1993)
show that about 70 per cent of the sky, as seen by the central source, 
has to be covered in order to produce the Fe K$\alpha$ line, 
if the broad line clouds have column densities 
of about 10$^{23}$ cm$^{-2}$, while the typical values for the 
BLR clouds covering fractions are of the order of 10--25 per cent 
(Davidson \& Netzer 1979; Goad \& Koratkar 1998). 
Alternatively, the Fe K$\alpha$ line may be produced by reflection by
the outer part of the accretion disc.

\subsection{Ionized Fe emission lines}

The spectrum of Mrk~509 shows emission from ionized iron,
consistent with either Fe XXV or Fe XXVI, implying photoionized gas 
outflowing or inflowing respectively. 
Alternatively, the ionized Fe K emission may be produced by reflection 
from the inner part of the accretion disc.

In fact, both the \xmm\ and the \suzaku\ data are consistent with the 
two scenarios, even if a slightly better fit ($\Delta\chi^2$=5.5 and 
9.7 for \xmm\ and \suzaku, respectively) is obtained in the case of broad line.
Moreover in the case of narrow emission lines the emitting gas should 
have a significant outflow (for Fe XXV, v$\sim$3500 and 2600 km s$^{-1}$ 
for \xmm\ and \suzaku, respectively) or inflow (for Fe XXVI, v$\sim$4500 
km s$^{-1}$) with velocities higher than what 
generally observed (Reynolds et al. 2004; Longinotti et al. 2007; 
but see also Bianchi et al. 2008 that detect an outflow of 
v=900$^{+500}_{-700}$ km s$^{-1}$). 
On the other hand, the high radiative efficiency of the 
source ($\eta$=0.12; Woo \& Urry 2002) suggests that the accretion 
disc is stable down to the innermost regions around the BH, 
where the reflection component should be shaped by relativistic effects. 
For these reasons, although an outflowing emitting gas is not excluded, 
the broad line interpretation seems favoured. 
In fact, the profile of the line is compatible with being shaped 
by relativistic effects, consistent with its origin being in 
the surface of an accretion disc, in vicinity of a black hole. 
Nevertheless, the width of the line is not a compelling evidence. 
The observed broadening of the line can be reproduced also 
with the Comptonization process occurring in the upper layer of the 
ionized accretion disc.
Moreover, we stress that the main evidences for the presence 
of a broad Fe K line comes from the mean summed spectrum.
The process of summing spectra, although is a powerful way to extract 
information, might be dangerous in presence of spectral variability 
and when applied to observations taken many years apart.  
Thus, the final answer on the origin of these ionized lines 
will be obtained with either a higher resolution observation 
or with significantly longer \xmm\ exposures.

\subsection{Ionized Fe absorption lines}

The \xmm\ data indicate the presence of a highly ionized 
absorption component, the best fit column density being
N$_H$=5.8$^{+5.2}_{-4.8}\times$10$^{22}$ cm$^{-2}$ and
ionization log($\xi$)=5.15$^{+1.25}_{-0.52}$. Moreover, fitting the absorption with
this model, it results that the absorber has to be blueshifted by
0.0484$^{+0.012}_{-0.013}$ c.  The blueshift corresponds to
an outflow velocity of $\sim$14000 km s$^{-1}$.  
The structure implies a significant blueshift if the
absorber is located in the core of Mrk~509 but, considering the
systemic velocity of the galaxy, its energy is also consistent with a
local absorber (McKernan et al. 2004; 2005; Risaliti et al. 2005; 
Young et al. 2005; Miniutti et al. 2007; but see also Reeves et al. 2008).
Nevertheless, the observed variability between the \xmm\ and
\suzaku\ observations points towards an origin within Mrk~509.

An hint of variability is observed around 6.7 keV both in the \xmm\ data
and by comparing the \xmm\ and \suzaku\ spectra. This could be due in
principle to variability in the red wing of the ionized emission
line. However, the total rms spectrum shows a peak of variability 
that is consistent with being narrow, thus it may suggest an 
alternative explanation. Indeed, the observed difference between 
the \xmm\ and \suzaku\ Fe K line shapes could be due to a further 
ionized absorption component, present only the \xmm\ observations,
with a column density N$_H$=5.4$^{+4.8}_{-4.4}\times$10$^{21}$
cm$^{-2}$ and ionization parameter log($\xi$)=2.04$^{+0.43}_{-0.60}$.  
When the structure at 6.7 keV is fitted with such a component, an absorption
structure appears around 7.3 keV, nevertheless its equivalent width is
not strong enough to reproduce the total absorption feature; moreover, 
it appears at slightly different energy, not completely fitting 
the $\sim$7.3 keV line. Thus, the absorption structures at 6.7 and 
the one at 7.3 keV may be connected and they may be indicative 
of another absorption screen. 
If this further lower ionization absorption component is present,
different absorption feature would be expected (due to the low 
ionization and high column density) at lower energies.
Smith et al. (2007) analyzed the RGS data and detected two absorption 
components with physical parameters similar 
(log($\xi$)=2.14$^{+0.19}_{-0.12}$ and 3.26$^{+0.18}_{-0.27}$; 
N$_H$=0.75$^{+0.19}_{-0.11}$ and 5.5$^{+1.3}_{-1.4}\times10^{21}$ 
cm$^{-2}$) to the ones that we infer, strengthening this interpretation.
There is also evidence for another, higher ionization, mildly 
relativistic, and variable ionized component in the XMM data. The study
of this more extreme component is addressed in another paper (Cappi et 
al., in preparation).

The observation of highly ionized matter in the core of Mrk~509 
is in line with its high BH mass and accretion rate.
In fact, we remind that at the Eddington limit the radiation pressure equals 
the gravitational pull, however the densities of the matter lowers with 
the BH mass (Shakura \& Sunyaev 1976). Thus the ionization of the material 
surrounding high accretion rate and BH mass AGNs, such as Mrk~509, should 
be higher than normal.
We stress, however, that in order to detail the physical 
parameters of the ionised emitter/absorber, further long observations 
are required.

\section{Conclusions}

The Fe K band of Mrk~509 shows a rich variety of emission/absorption 
components. The \xmm\ and \suzaku\ data shows  evidence for the
presence of: 
\begin{itemize}
\item a resolved, although not very broad, ($\sigma\sim$0.07 keV)
neutral Fe K$\alpha$ line and associated Fe K$\beta$ emission.
The width of the line suggests that the 6.4 keV line is produced 
in the outer part of the accretion disc (the broad line region or torus 
emission seem unlikely). 
The measured reflection fraction is consistent in this case with 
the intensity of the line, while a covering factor or column density 
higher than generally observed would be required if the line were 
produced in the BLR or the torus;
\item both the \suzaku\ and the \xmm\ data show an excess 
due to ionized Fe K emission. Both datasets show a superior 
fit when a broad ionized line coming from the central parts 
of the accretion disc is considered. 
The data are inconsistent with narrow emission from
a distant scattering material at rest, while it can not be excluded 
if the gas is outflowing (v$\sim$3500 km s$^{-1}$)
\item both EPIC--pn and MOS data show an absorption line 
at $\sim$7.3 keV, present in the summed spectrum of all \xmm\ 
observations only. This component confirms the presence of highly ionized, 
outflowing (v$\sim$14000 s$^{-1}$), gas along the line of sight. 
The comparison between \xmm\ and \suzaku\ suggests a variability of this 
component;
\item a hint of an enhancement of variability~-~both by
considering the \xmm\ data alone and by comparison between the two data
sets~-~at $\sim$6.7 keV that could be either due to the high
variability of the red wing of the broad ionized Fe K line, 
possibly associated with a variation of the ionisation of the disc, 
or to a second ionized absorption line.
\end{itemize}

\section*{Acknowledgments}

This paper is based on observations obtained with \xmm, an ESA science 
mission with instruments and contributions directly funded by 
ESA Member States and NASA. This work was partly supported by 
the ANR under grant number ANR-06-JCJC-0047.
GP, CV and SB thank for support the Italian Space Agency 
(contracts ASI--INAF I/023/05/0 and ASI I/088/06/0). 
GM acknowledge funding from Ministerio de Ciencia e Innovaci\'on 
through a Ram\'on y Cajal contract. GP thanks Regis Terrier, 
Andrea Goldwurm and Fabio Mattana for useful discussion.
We would like to thank the anonymous referee for the detailed 
reading of the manuscript and for the comments that greatly 
improved the readability of the paper.

\end{document}